\PassOptionsToPackage{unicode}{hyperref}
\PassOptionsToPackage{hyphens}{url}

\PassOptionsToPackage{table}{xcolor}

\documentclass[
  10pt,
  letterpaper,
]{article}
\usepackage{xcolor}
\usepackage[top=0.85in,left=2.75in,footskip=0.75in]{geometry}
\usepackage{amsmath,amssymb}
\usepackage{iftex}
\ifPDFTeX
  \usepackage[T1]{fontenc}
  \usepackage[utf8]{inputenc}
  \usepackage{textcomp} 
\else 
  \usepackage{unicode-math} 
  \defaultfontfeatures{Scale=MatchLowercase}
  \defaultfontfeatures[\rmfamily]{Ligatures=TeX,Scale=1}
\fi
\usepackage{lmodern}
\ifPDFTeX\else
\fi
\IfFileExists{upquote.sty}{\usepackage{upquote}}{}
\IfFileExists{microtype.sty}{
  \usepackage[protrusion=true,expansion=false]{microtype}
  \UseMicrotypeSet[protrusion]{basicmath} 
}{}
\makeatletter
\@ifundefined{KOMAClassName}{
  \IfFileExists{parskip.sty}{%
    \usepackage{parskip}
  }{
    \setlength{\parindent}{0pt}
    \setlength{\parskip}{6pt plus 2pt minus 1pt}}
}{
  \KOMAoptions{parskip=half}}
\makeatother

\usepackage{longtable,booktabs,array}
\usepackage{calc} 
\usepackage{etoolbox}
\makeatletter
\patchcmd\longtable{\par}{\if@noskipsec\mbox{}\fi\par}{}{}
\makeatother
\IfFileExists{footnotehyper.sty}{\usepackage{footnotehyper}}{\usepackage{footnote}}
\makesavenoteenv{longtable}
\usepackage{graphicx}
\makeatletter
\newsavebox\pandoc@box
\newcommand*\pandocbounded[1]{
  \sbox\pandoc@box{#1}%
  \Gscale@div\@tempa{\textheight}{\dimexpr\ht\pandoc@box+\dp\pandoc@box\relax}%
  \Gscale@div\@tempb{\linewidth}{\wd\pandoc@box}%
  \ifdim\@tempb\p@<\@tempa\p@\let\@tempa\@tempb\fi
  \ifdim\@tempa\p@<\p@\scalebox{\@tempa}{\usebox\pandoc@box}%
  \else\usebox{\pandoc@box}%
  \fi%
}
\def\fps@figure{htbp}
\makeatother

\NewDocumentCommand\citeproctext{}{}

\makeatletter
 \let\@cite@ofmt\@firstofone
 \def\@biblabel#1{}
 \def\@cite#1#2{{#1\if@tempswa , #2\fi}}
\makeatother
\newlength{\cslhangindent}
\newlength{\csllabelwidth}
\newenvironment{CSLReferences}[2] 
 {\begin{list}{}{%
  \setlength{\itemindent}{0pt}
  \setlength{\leftmargin}{0pt}
  \setlength{\parsep}{0pt}
  \ifodd #1
   \setlength{\leftmargin}{\cslhangindent}
   \setlength{\itemindent}{-1\cslhangindent}
  \fi
  \setlength{\itemsep}{#2\baselineskip}}}
 {\end{list}}
\usepackage{calc}

\newcommand{\CSLLeftMargin}[1]{\parbox[t]{\csllabelwidth}{\strut#1\strut}}
\newcommand{\CSLRightInline}[1]{\parbox[t]{\linewidth - \csllabelwidth}{\strut#1\strut}}

\usepackage{changepage}

\usepackage{marvosym}

\usepackage{nameref,hyperref}

\ifx\XeTeXversion\@undefined
  \DisableLigatures[f]{encoding = *, family = * }
\fi

\newcolumntype{+}{!{\vrule width 2pt}}

\newlength\savedwidth

\raggedright
\usepackage[aboveskip=1pt,labelfont=bf,labelsep=period,justification=raggedright,singlelinecheck=off]{caption}
\renewcommand{\figurename}{Fig}

\makeatletter
\renewcommand{\@biblabel}[1]{\quad#1.}
\makeatother

\usepackage{lastpage,fancyhdr}
\usepackage{epstopdf}
\fancyheadoffset[L]{2.25in}
\fancyfootoffset[L]{2.25in}
\usepackage{setspace}
\microtypesetup{protrusion=true, expansion=false}
\makeatletter
\@ifpackageloaded{caption}{}{\usepackage{caption}}
\AtBeginDocument{%
\ifdefined\contentsname
  \renewcommand*\contentsname{Table of contents}
\else
  \newcommand\contentsname{Table of contents}
\fi
\ifdefined\listfigurename
  \renewcommand*\listfigurename{List of Figures}
\else
  \newcommand\listfigurename{List of Figures}
\fi
\ifdefined\listtablename
  \renewcommand*\listtablename{List of Tables}
\else
  \newcommand\listtablename{List of Tables}
\fi
\ifdefined\figurename
  \renewcommand*\figurename{Figure}
\else
  \newcommand\figurename{Figure}
\fi
\ifdefined\tablename
  \renewcommand*\tablename{Table}
\else
  \newcommand\tablename{Table}
\fi
}
\@ifpackageloaded{float}{}{\usepackage{float}}
\floatstyle{ruled}
\@ifundefined{c@chapter}{\newfloat{codelisting}{h}{lop}}{\newfloat{codelisting}{h}{lop}[chapter]}
\floatname{codelisting}{Listing}

\makeatother
\makeatletter
\@ifpackageloaded{caption}{}{\usepackage{caption}}
\@ifpackageloaded{subcaption}{}{\usepackage{subcaption}}
\makeatother
\usepackage{bookmark}
\IfFileExists{xurl.sty}{\usepackage{xurl}}{} 
\hypersetup{
  pdftitle={Ten simple rules for teaching data science},
  pdfauthor={Tiffany A. Timbers; Mine Çetinkaya-Rundel},
  hidelinks,
  pdfcreator={LaTeX via pandoc}}

\begin{document}
\vspace*{0.2in}

\begin{flushleft}
{\Large
\textbf\newline{Ten simple rules for teaching data
science} 
}
\newline
\\
Tiffany A. Timbers\textsuperscript{1*}, Mine
Çetinkaya-Rundel\textsuperscript{2}
\\
\bigskip
\textbf{1} Department of Statistics, The University of British Columbia,
Vancouver, BC, Canada, \\ \textbf{2} Department of Statistical Sciences,
Duke University, Durham, NC, USA, 
\bigskip

%
%
* tiffany.timbers@stat.ubc.ca (TT)

\end{flushleft}


\section{Introduction}\label{sec-intro}

Data science is the study, development, and practice of using
reproducible and transparent processes to generate insight from data
{[}1--4{]}. With roots in statistics and computer science, data science
educators draw on many of the teaching strategies used in those fields
{[}5--7{]}. However, data science is a distinct discipline with its own
unique challenges and opportunities for teaching and learning. Here, we
collate and present ten simple rules for teaching data science, piloted
by leading data science educators in the community and successfully
applied in our own data science classroom.

\section{Rule 1: Teach data science by doing data
analysis}\label{sec-rule1}

The first rule is teaching data science by doing data analysis. This
means that in your first data science lesson, not in your third lesson,
not in your 10th lesson, not at the end of the semester, but in your
first data science lesson, have the students load some data, perform
some simple data wrangling, and create a data visualization. We call
this the ``let them eat cake'' approach to teaching data science
{[}8{]}.

\begin{figure}

\begin{minipage}{\linewidth}
\pandocbounded{\includegraphics[keepaspectratio]{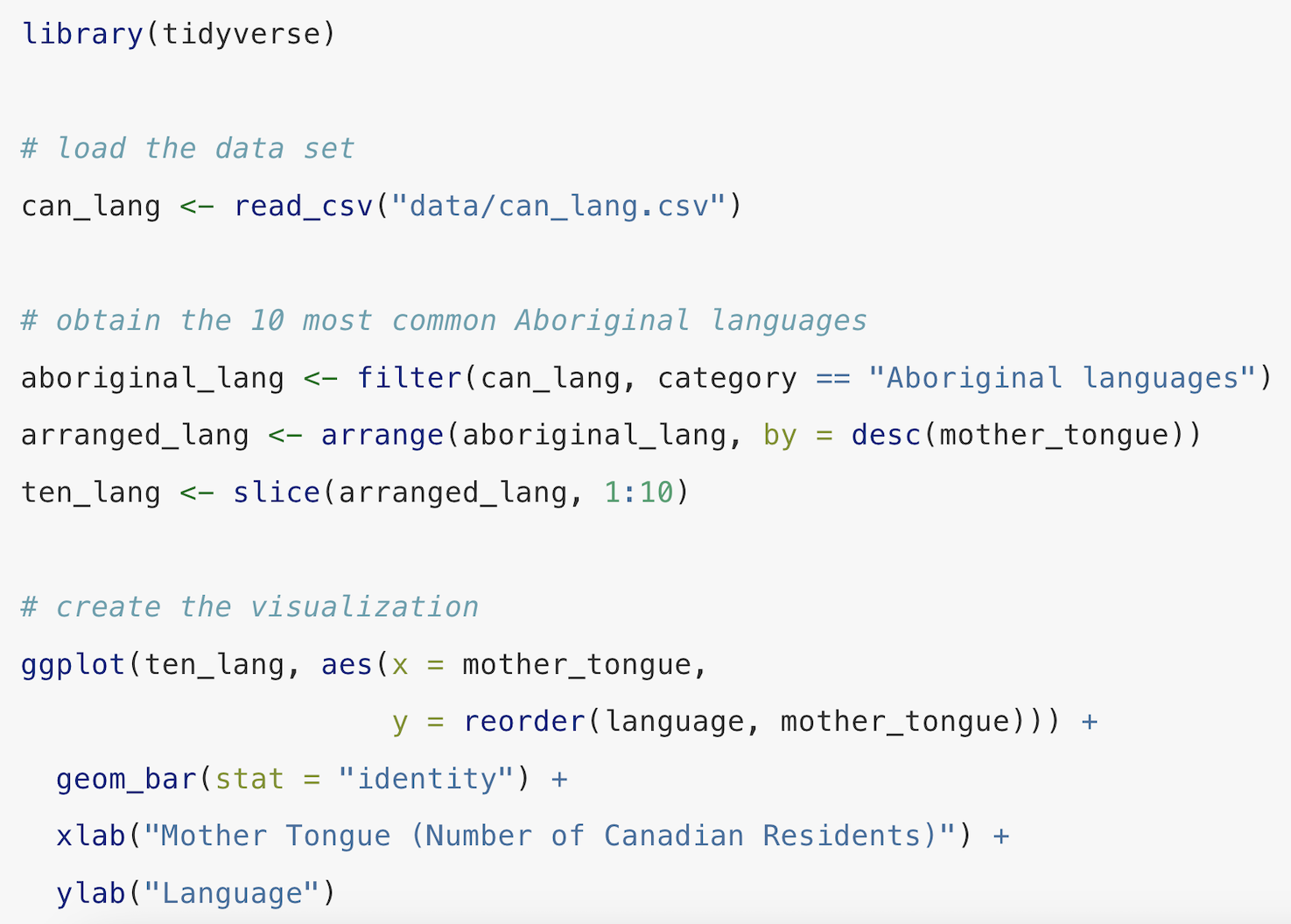}}\end{minipage}%
\newline
\begin{minipage}{\linewidth}

\begin{figure}[H]

\centering{

\includegraphics[width=1\linewidth,height=\textheight,keepaspectratio]{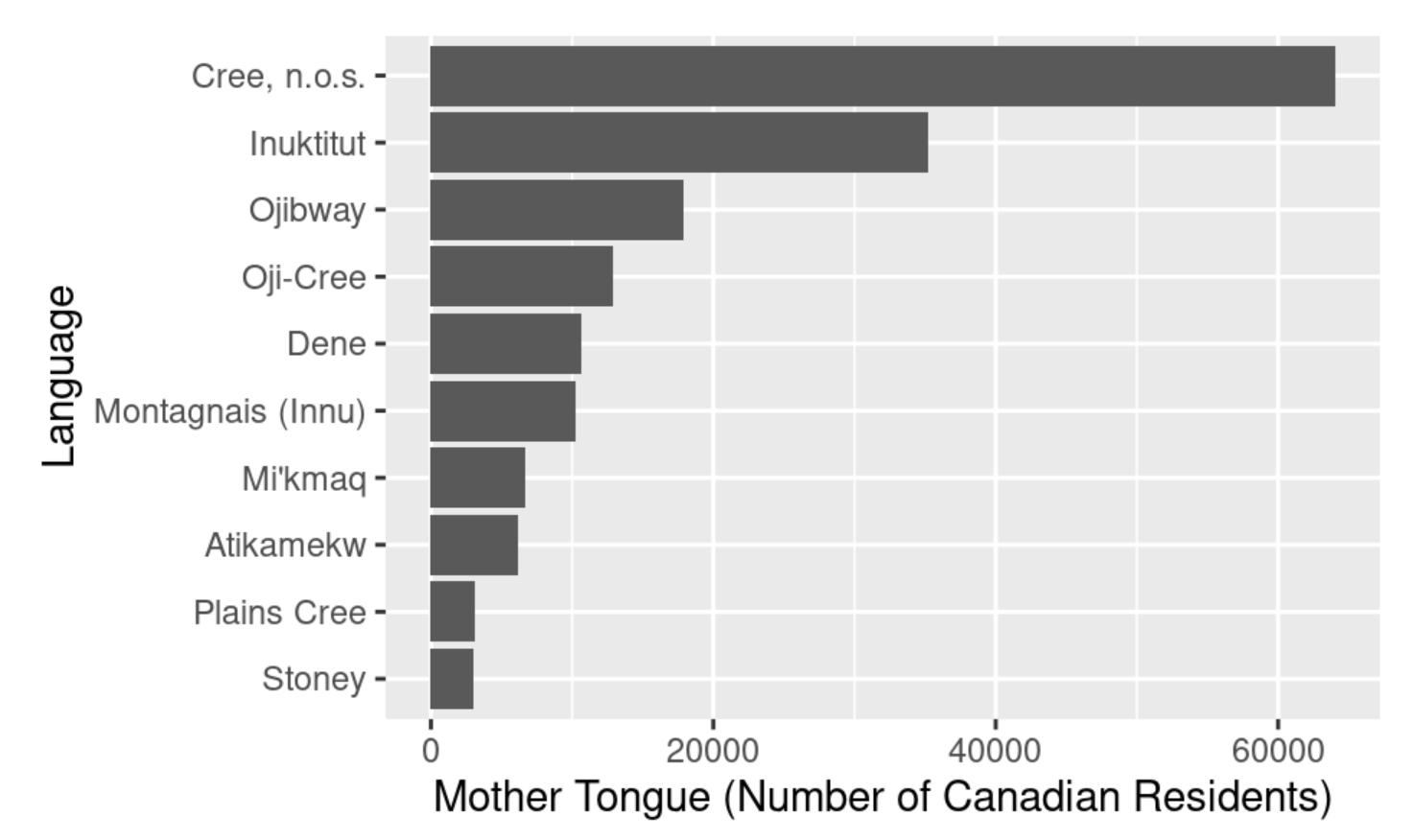}

}

\caption{\label{fig-intro-ds-code}Example code from the first chapter of
\emph{Data Science: A First Introduction} {[}4{]} that gets students
doing data analysis on day one.}

\end{figure}%

\end{minipage}%

\end{figure}

Why do we suggest this? Because, just like eating cake, it's extremely
motivating to students. At first, students sign up for a data science
course or workshop because they're interested in asking and answering
questions about the world using data. They likely don't yet have
sufficient knowledge to care deeply about detailed, technical aspects,
such as object data types, whether to use R versus Python, or, if using
R, whether to opt for the tidyverse or base R. As a result, we should
show them something interesting very early on to hook them. After they
are hooked, they will be begging you to answer questions about the
detailed, technical aspects you intentionally omitted. An example of
this is shown in Fig~\ref{fig-intro-ds-code}; code from the first
chapter of \emph{Data Science: A First Introduction} {[}4{]}. In this
first chapter, we ask learners to load data from a CSV file and perform
introductory data wrangling using filtering, arranging, and slicing.
Finally, create a plot to answer a question about indigenous languages
in Canada: how many people living in Canada speak an indigenous language
as their mother tongue? Other leading data science educators who
advocate and practice this rule include Wang and colleagues {[}9{]},
David Robinson in his introductory online Data Science course
{[}10,11{]}, and Jenny Bryan {[}12{]}, who summed this up nicely in a
tweet ``{[}\ldots{]} I REPEAT, do not front-load the `boring',
foundational stuff. Do realistic and non-boring tasks and work your way
down to it.''

\section{Rule 2: Use participatory live coding}\label{sec-rule2}

The second rule is to use participatory live coding. This means that
when you are working with code in the classroom, instead of showing it
on a static slide or just running it in an executable slide or an IDE,
actually type the code and narrate it as you teach. Have the
participants follow along as well. The reason is that it demonstrates
your best practices for processes and workflows, topics that are
important in practice but, unfortunately, are often just an afterthought
in teaching computational subjects. You can discuss why you're
approaching things differently as you go. You are also likely to make
mistakes as you live-code, and that's actually a good thing. It helps
you appear human to students because they, too, will make mistakes. More
importantly, it allows you to demonstrate how you approach debugging to
solve code problems, which they can leverage in their homework and later
in their work outside the course. Participatory live coding also slows
you down, so you don't go too fast for the students. This pedagogy
originates from the ``I do, we do, you do'' method of knowledge transfer
{[}13{]}, and its application in teaching programming was pioneered by
The Carpentries, a global nonprofit (\url{https://carpentries.org}).
Best practices for doing this have been refined and shared as ten quick
tips by Nederbragt and colleagues {[}14{]}.

\section{Rule 3: Give tons and tons of practice and timely
feedback}\label{sec-rule3}

The third simple rule is to give tons and tons of practice. Give the
learners many, many, many problems to solve, probably many, many, many
more than you think they might need. The reason for this is that
repetition leads to learning {[}15{]}. That's not just in humans; it is
more fundamental than that. In the field of animal behavior across the
animal kingdom, repetition is found to lead to learning {[}16,17{]}.
Similarly, students need to practice tasks many times to understand and
then perform them effectively. For example, when teaching students to
read data from a file, don't just give them one file to read; instead,
have them work with six different variants of a very similar file. With
this approach, students have to investigate each file in detail,
including checking the file type, column spacing, whether there's
metadata to skip, whether there are column names, etc. In our courses,
they will complete these six variants in an in-class worksheet, a lab
assignment, and a quiz. Meaning that by the end of the course, they will
have practiced this skill over 15 times. Many excellent data science
educational resources use this pedagogy, including software packages
(e.g., the swirl R package {[}18{]}), online courses (e.g., Kaggle Learn
{[}19{]}), and popular textbooks (e.g., R for Data Science {[}20{]}).
For those new to designing practice exercises for data science, we
recommend looking at the ``Exercise Types'' chapter of \emph{Teaching
Tech Together: How to make your lessons work and build a teaching
community around them} by Greg Wilson {[}21{]}.

\begin{figure}

\centering{

\includegraphics[width=1\linewidth,height=\textheight,keepaspectratio]{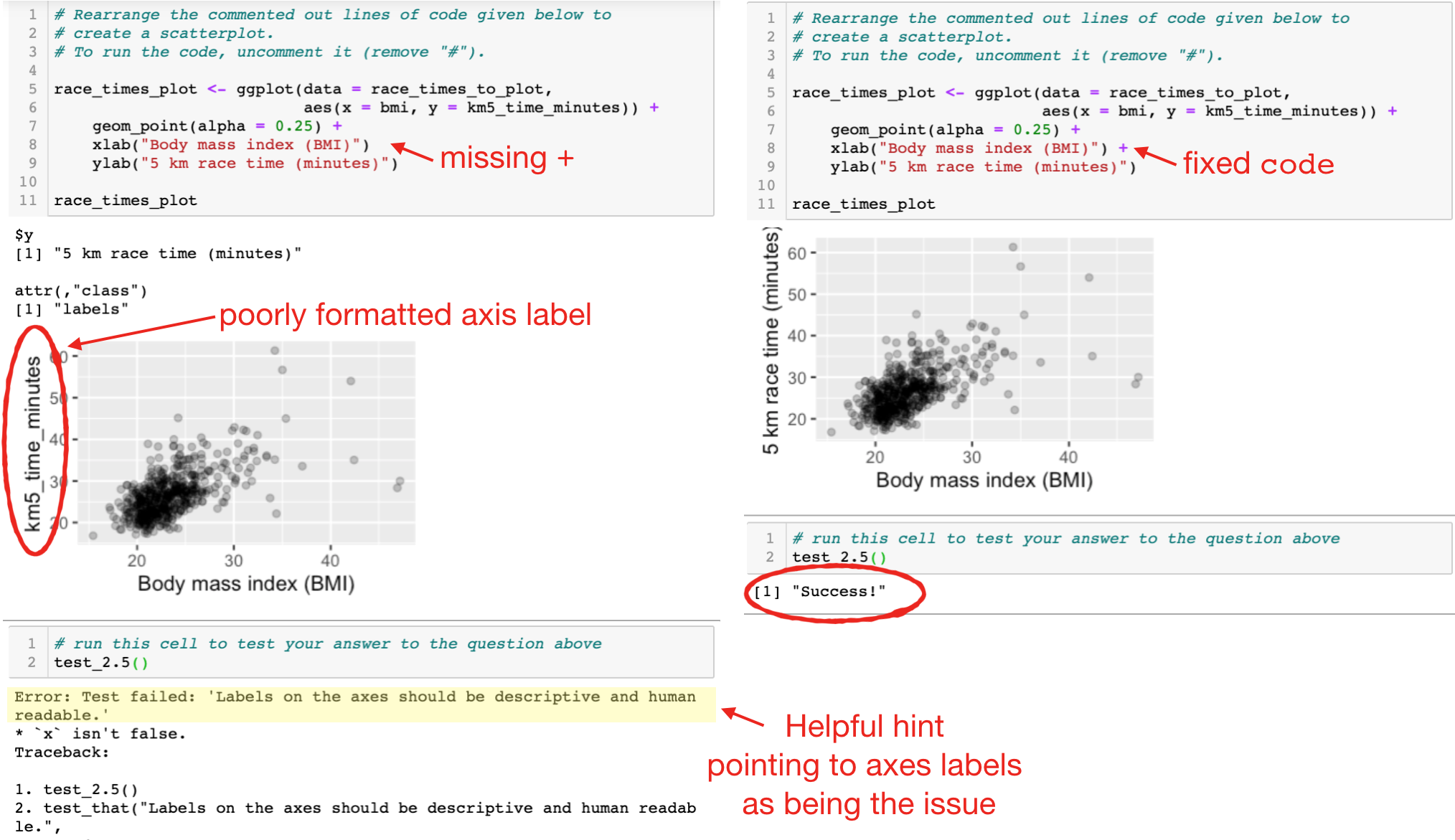}

}

\caption{\label{fig-code-feedback}Example of automated software test
feedback to students.}

\end{figure}

When giving lots of practice, pair it with timely feedback. Practice
without feedback has limited value. So, how can we provide a wealth of
timely feedback, especially with our limited teaching capacity and
resources? One way we can do this is through automated software tests.
In data science, many of the problems we assign to students involve
writing code. As a consequence, we can write software tests that provide
feedback to students, letting them know when they give a wrong answer in
a particular way, as well as providing a gentle, helpful nudge to solve
the problem in a different way. Fig~\ref{fig-code-feedback} shows an
example of this in practice. Here, students were given some ggplot code
in a Parsons problem format (the lines of code were given in the wrong
order, and the students needed to rearrange them to the correct order).
In this example, a student has rearranged the code, but while doing so
has introduced a syntax error. As a result, a plot is created, but it's
not quite what we expect. Without timely feedback, the student might not
realize that there's a problem with their code until much later, or not
at all if they fail to check the feedback and solutions after the
assignment is graded and grades are returned (often days or weeks
later). Automated software testing can provide timely feedback while
students' focus and attention are on the topic being learned and
practiced. This pedagogy was first developed and used for teaching
programming in computer science courses {[}reviewed in 21{]} and is now
being adopted in data science courses. There are now many wonderful and
popular software packages for this in the context of data science,
including the learnr R package {[}22{]} for R code, and the NBgrader
{[}23{]} and Otter Grader {[}24{]} software packages, which work for
both R and Python code.

\section{Rule 4: Use tractable or toy data examples}\label{sec-rule4}

Our fourth simple rule is to use tractable or toy data examples when
introducing a new tool, method, or algorithm to students. Tractable or
toy data sets have a countable number of elements, which can fit in our
working memory. This allows students to track the progression of
everything through the algorithm's steps, see how the elements are
manipulated, and gain a deeper understanding of these concepts. For
example, in one of our courses, we use the Palmer penguins data set
{[}25{]} to introduce the students to k-means clustering. Instead of
giving them the entire data set, which contains hundreds of
observations, we first subset it to just a handful. Then we can walk the
students through what happens to these observations at each step of the
algorithm, building understanding and intuition for the algorithm.
Inspiration from this comes from Jenny Bryan's great dplyr joins cheat
sheet {[}26{]}. In her cheat sheet, she teaches all the various joins
from the dplyr package {[}27{]}. This is a difficult topic for students
to understand and memorize, and when instructors teach this with large
data sets, it is really hard for students to get an idea of what's going
on. To make matters worse, all of these joins are also similarly named
(e.g, left join, right join, inner join, outer join, etc). To address
this issue, Jenny Bryan's cheat sheet uses two toy data sets on
superhero comic characters and publishers. The cheat sheet outlines all
the possible joins, narrating and displaying the output of each. The
superhero dataset has only 7 rows and 4 columns, while the publisher
dataset has 3 rows and 2 columns. These data sets are small enough to
fit in learners working memory, making the joins tractable and much
easier to understand.

It is, of course, not feasible to keep learners interested and motivated
to learn more if we rely solely on toy datasets. So, once students have
a conceptual understanding of the tool, method, or algorithm being
taught, we can move on to using real, rich datasets, which is our next
rule.

\section{Rule 5: Use real and rich, but accessible data
sets}\label{sec-rule5}

After you've helped students develop a conceptual understanding of the
new tool, method, or algorithm you are teaching, the next step is to
have students apply it to realistic questions and real, rich data.
However, as you're doing this, it is critical to ensure that the dataset
is also accessible to all your learners. The question, as well as the
observations (i.e., rows) and variables (i.e., columns) in the dataset,
must be things that your learners can quickly understand. This can
easily become an expert blind spot for us when we are teaching,
especially if we have training in a particular domain. For example, one
of the authors, who is trained in the biological sciences, might think
that using a deep sequencing data set would be a great and motivating
example for a particular algorithm we want to teach students. However,
this thought likely stems from the author's deeper understanding of
biological processes, which, in this example, gives them an expert blind
spot. Given that, if learners do not have similar background knowledge,
that dataset might not be appropriate. It might introduce too much
cognitive load for the students, to the extent that they cannot focus on
the task at hand --- refining their understanding and application of the
tool, method, or algorithm being taught. We do not want to use up our
students' limited cognitive resources to understand what the dataset is
about. Instead, we want to use something that all our learners can
easily understand, including what the observations are and what the
columns represent. One example from one of the courses that we teach
uses Canadian census data about the languages spoken in Canada across
different regions {[}28{]}. Other examples include the Gapminder
{[}29{]} and UN Votes {[}30{]} data sets. These are particularly nice
examples because they contain hundreds of observations; however, the
observations are something most people can understand: a country, a
year, a population, a vote, etc. And with such a dataset, we can ask
questions that most learners are interested in, because everyone has
grown up in at least one country, and possibly more, in their lifetime.
And we all grew up at different times in history. These lived
experiences make us generally knowledgeable and interested in asking
questions about the datasets. There are many more real and rich data
sets that are accessible and usable. The main point here is that it
shouldn't take your students long to understand the data set, because a
deep understanding of it isn't what you are trying to teach at that
moment. It can also be helpful to bring in data sets that reflect your
learners' local context (e.g., country, region, culture) and/or current
events to increase engagement and relevance. It can be difficult to find
such data sets in a timely manner, but it's often worth the effort for
learners to connect what they are learning in their data science course
to what's happening in their own lives and communities outside the
classroom.

\section{Rule 6: Provide cultural and historical
context}\label{sec-rule6}

Our sixth simple rule is to provide a cultural and historical context
for what you're teaching. For example, when teaching how to use a new
software tool or a new feature of a tool that students already know how
to use, and things do not seem optimally designed from the learner's
perspective, it's really helpful to explain why they are that way. If
you give the design and historical context, for example, saying people
thought about this when they built the tool and decided this was the
best way to implement it for reasons X, Y, and Z, it helps the students
understand that software tools are built by humans, and so they are
going to be influenced by humans' perspective, human history, and human
culture. Furthermore, it helps prevent frustration or annoyance with the
software because they can rationalize why it works the way it does. We
believe it is crucial to help prevent these frustrations or annoyances,
as we have observed that, for some learners, they can become significant
barriers and lead to a dislike or avoidance of a particular piece of
software.

For example, when we teach the programming language R {[}31{]} using the
suite of tidyverse R packages {[}32{]}. Learners observe that these
packages heavily use unquoted column names when referencing data frame
columns in their function calls. That is really strange when you come
from other programming languages, as most other programming languages
require quoted strings when referencing object attributes (which is what
a data frame column is). This can also make writing functions that
utilize tidyverse functions a bit more challenging due to issues with
indirection. For learners with experience in Python, Java, or C, this
might initially seem like a really bad idea. However, once they learn
that R and the suite of tidyverse R packages were written by
statisticians for performing data analysis and graphics {[}31,32{]}, and
that they designed the language and packages with the expectation that
much of the users time would be frequently typing things into the
console or running code interactively, it makes a lot more sense that
they would want to minimize the amount of typing and tracking of opening
and closing quotations for their users. Doing this minimizes the
potential syntax errors that could be introduced by errors in forgetting
to open or close quotations.

Another example comes from the use of the arrow (\textless-) as an
assignment operator in R, which may seem odd to some learners as it uses
two characters. Again, however, when given the historical context that R
was derived from another programming language named S, and S was
inspired by another programming language, APL, that was designed for a
particular keyboard with one key mapped to the assignment operator
{[}33{]}, it makes a lot more sense as to why that design choice was
made. Design and historical contexts help learners understand the
rationale behind different design choices, allowing them to see choices
they might not initially agree with as excellent within these contexts.
As an aside, for those interested in learning more about the history of
the R programming language, see the ``History and Overview of R''
chapter in Roger Peng's \emph{R programming data science} book {[}34{]}.

\section{Rule 7: Build a safe, inclusive, and welcoming
community}\label{sec-rule7}

The next rule is to build a safe, inclusive, and welcoming community. If
we were to rewrite this paper, we would probably move this to rule
number one, given that this is the first thing you need to do when
teaching any subject. The reason is that people don't learn effectively
when they don't feel psychologically safe. Psychological safety is the
belief that one can express oneself, through speech or actions, without
fear of negative consequences or feedback {[}35{]}. If learners do not
feel safe asking questions without being made to look or feel dumb, they
are not going to ask questions, and they are not going to be engaged
{[}36{]}. If they don't feel safe from negative perceptions about their
intelligence, discrimination, or harassment in the classroom (or related
spaces such as office hours, course forums, study groups, etc.), they
may become so disengaged that they even stop showing up. Thus, creating
safe, inclusive, and welcoming learning environments is crucial for
effective learning, and as instructors, we have a responsibility to
establish scaffolding and guidelines that facilitate this. One thing we
do in our courses is establish a course code of conduct, which holds a
place of prevalence in the classroom and related learning spaces. At the
beginning of a new course, we take time in the first class to present
our code of conduct to the learners. The codes of conduct we use are
very explicit. They discuss expected behaviors, behaviors that will not
be tolerated, the process for reporting violations, and the consequences
for violating these rules. We also ensure that there are multiple ways
to report violations, to support students in the unlikely event that the
instructor is the one violating the code of conduct. It is also
important that when a violation is reported, it be taken seriously and
acted upon in a timely manner. No student concern should ever be ignored
or brushed off. For instructors seeking to establish a code of conduct
for their course, we recommend examining existing codes of conduct from
other organizations and tailoring them to your specific context. One
particularly good example is The Carpentries' code of conduct
(\url{https://docs.carpentries.org/topic_folders/policies/code-of-conduct.html}).

\section{Rule 8: Use checklists to focus and facilitate peer
learning}\label{sec-rule8}

Rule number eight is to use checklists to focus and facilitate peer
learning. A well-documented pedagogical best practice is for peers to
learn from each other {[}37{]}. One implementation of this is peer
review. However, peer review can be particularly challenging if you
haven't done it before, or if it involves reviewing something new that
you are still learning. So what can we do as instructors to facilitate
this practice for our learners? Given that we are well-practiced at
creating rubrics for grading student work, we can use our existing
rubrics as a starting point for drafting peer-review checklists. Why
checklists? Checklists can help ensure complex tasks are completed
successfully and have been used in safety-critical systems (such as
aviation, surgery, or nuclear power). They can be particularly helpful
for complex tasks, as well as for repetitive ones that are consequently
boring {[}38{]}. For these reasons, checklists have recently been
adopted in scientific (e.g., PLoS journals, Nature Ecology \& Evolution,
Journal of Open Source Education, Journal of Open Source Software) and
software (e.g., ROpenSci, PyOpenSci) publishing to help ensure that
reviewers and editors increase transparency, decreases bias and call
attention to essential elements of reviews that are often overlooked
{[}39{]}. As reviewers for some of these journals and organizations, the
authors have found checklists extremely helpful for focusing our
attention on the most important elements of a review and for ensuring we
don't miss anything. We believe this idea also has value in training
data science students. \textbf{S1 File} shows an example of a data
analysis review checklist that we have used in our data science courses.
It serves to communicate the aspects that we, as educators, believe are
important for the assessment to be completed to a high quality. In
addition to checking off the checklist items, students are also asked to
provide written comments and feedback. This checklist helps focus
students' comments and feedback on the items they were unable to check
off. For example, if there were issues with the software tests (e.g.,
they were missing) and issues with the discussion section (e.g., they
did not mention any limitations of the analysis), they would not check
those boxes, and that would help focus their review comments on critical
feedback about the issues with those sections.

\section{Rule 9: Teach students to work
collaboratively}\label{sec-rule9}

Data science is a highly collaborative discipline. Data scientists often
work in teams or, at a minimum, need to collaborate with other
stakeholders (e.g., domain experts, project managers, clients).
Consequently, it is essential that we teach our students to work
collaboratively. To do this effectively, we must teach them both the
technical tools and skills necessary for collaboration (e.g., version
control tools, such as Git and GitHub, project boards), as well as the
social practices that make collaboration effective (e.g., active
listening, giving critical feedback, code review). We must also give
them opportunities to practice collaboration. Practice should occur at
increasing levels of complexity, starting with smaller in-class
activities, such as think-pair-share and pair programming, progressing
to larger group assignments, and then culminating in projects.

When getting students to work collaboratively on longer-scale
assignments and projects, it is important to scaffold good collaboration
practices into your assignment or project expectations (and grading). In
our project courses, we have students spend the first working session
almost entirely on group formation activities. In this session, we have
them participate in icebreaker activities to get to know each other and
build trust. After that, we get them to create a teamwork contract that
outlines their expectations for the project and how they will work
together. During the project, we have them hold regular team check-ins
(e.g., stand-ups, meetings) and use project boards to track their tasks
and progress. At the end of the project, we have them do a teamwork
reflection activity to reflect on what went well and what could be
improved. As seasoned collaborators, we often take these practices for
granted. However, for students new to collaboration, these practices are
often missed, and as a result, collaboration frequently breaks down.
Even with these scaffolds in place, we still see some groups struggle
with collaboration. To manage this, we recommend discussing with
students the importance of expecting, rather than avoiding, conflict,
and developing, as a group, a plan ahead of time for how to manage it
when it arises.

\section{Rule 10: Have students do projects}\label{sec-rule10}

Our final simple rule is to have students complete projects --- create
an entire data product (ideally, on a topic of their choosing, if
feasible) from beginning to end (or from ``nachos to cheesecake,'' to
quote Jenny Bryan). Projects can be done individually or in groups. The
main point is that students have the opportunity to experience the
entire data science workflow. Doing project work is important because it
really helps motivate students. It also provides students with valuable
experience in dealing with the messiness of real data (we all know that
in the real world, data is often quite messy). Additionally, courses
typically focus on just a small part of a data analysis workflow. For
example, they might focus on data visualization, data wrangling, or
modeling. But in a project, they get to see how all these pieces fit
together in a more realistic scenario. This is critical training for any
aspiring data scientist {[}40{]}.

From the instructor's perspective, however, projects can feel daunting,
particularly when student numbers are large and teaching resources are
not. One way to make projects more feasible is to scope them. You may
want to limit the project to a particular topic, or have students choose
a data set from a given list, use a specific method, or employ a
particular programming language. Doing this provides some consistency in
grading and allows you to create a rubric that applies to all projects.
An example of a scoped project from a course on collaborative software
development in data science that we teach asks students to create a
Python package with n functions, where n is the number of students in
the project group. The functions must be related to a common theme and
fall under the umbrella of data science. Students must utilize the
packaging tools and collaborative practices taught in our course. This
kind of project is very feasible to grade because its scope is
well-defined and limited. However, it also allows students to be
creative and work on something they are interested in. Another way to
scope projects is to have students work on projects that use data
related to their own research or thesis work, but utilize the data
science concepts, tools or techniques being taught in the course. This
is particularly relevant for graduate students enrolled in data science
courses as part of their degree program. For example, in the UBC STAT
545 course developed by Jenny Bryan, students learn new data science
concepts and skills using a tractable dataset in the classroom (e.g.,
the Gapminder dataset), but their project involves applying the newly
acquired concepts and skills to their own research or thesis work. The
project complexity builds week-by-week along with what is being taught
in the course syllabus. Again, this approach makes grading more feasible
by creating homogeneity with what data science techniques and methods
the students use in their projects, and students get to work on
something they are deeply interested in and motivated to do.

\section{Conclusion}\label{sec-conclusion}

This list of ten simple rules for teaching data science is by no means
exhaustive, but we hope it provides a useful starting point for new data
science educators. This list was curated from our own experiences
teaching data science, as well as from what we've seen other leading
data science educators practice.

\section{Supporting information}\label{supporting-information}

\paragraph*{S1 File.}
\label{s1-file}
{\textbf{S1\_File.pdf. Example data analysis peer review checklist used
in data science courses taught by the authors.}}

\section{Acknowledgements}\label{acknowledgements}

Many thanks to the leading data science educators in our community who
pioneered the teaching practices outlined in our ten simple rules and
the students with whom we have tested them on in our own classrooms.

\section*{References}\label{references}
\addcontentsline{toc}{section}{References}

\phantomsection\label{refs}



\begin{CSLReferences}{0}{1}
\bibitem[\citeproctext]{ref-berman2018realizing}
\CSLLeftMargin{1. }%
\CSLRightInline{Berman F, Rutenbar R, Hailpern B, Christensen H,
Davidson S, Estrin D, et al. Realizing the potential of data science.
Communications of the ACM. 2018;61: 67--72. }

\bibitem[\citeproctext]{ref-Wing2019Data}
\CSLLeftMargin{2. }%
\CSLRightInline{Wing JM. The {Data} {Life} {Cycle}. Harvard Data Science
Review. 2019;1.
doi:\href{https://doi.org/10.1162/99608f92.e26845b4}{10.1162/99608f92.e26845b4}}

\bibitem[\citeproctext]{ref-Irizarry2020Role}
\CSLLeftMargin{3. }%
\CSLRightInline{Irizarry RA. {The Role of Academia in Data Science
Education}. Harvard Data Science Review. 2020;2.
doi:\href{https://doi.org/10.1162/99608f92.dd363929}{10.1162/99608f92.dd363929}}

\bibitem[\citeproctext]{ref-timbers2022data}
\CSLLeftMargin{4. }%
\CSLRightInline{Timbers T, Campbell T, Lee M. Data science: A first
introduction. Chapman; Hall/CRC; 2022. }

\bibitem[\citeproctext]{ref-carver2016guidelines}
\CSLLeftMargin{5. }%
\CSLRightInline{Carver R, Everson M, Gabrosek J, Horton N, Lock R, Mocko
M, et al. Guidelines for assessment and instruction in statistics
education (GAISE) college report 2016. AMSTAT; 2016. }

\bibitem[\citeproctext]{ref-zendler2015}
\CSLLeftMargin{6. }%
\CSLRightInline{Zendler A, Klaudt D. Instructional methods to computer
science education as investigated by computer science teachers. Journal
of Computer Science. 2015;11: 915--927.
doi:\href{https://doi.org/10.3844/jcssp.2015.915.927}{10.3844/jcssp.2015.915.927}}

\bibitem[\citeproctext]{ref-fincher2019cambridge}
\CSLLeftMargin{7. }%
\CSLRightInline{Fincher SA, Robins AV. The cambridge handbook of
computing education research. Cambridge University Press; 2019. }

\bibitem[\citeproctext]{ref-ccetinkaya2021fresh}
\CSLLeftMargin{8. }%
\CSLRightInline{Çetinkaya-Rundel M, Ellison V. A fresh look at
introductory data science. Journal of Statistics and Data Science
Education. 2021;29: S16--S26. }

\bibitem[\citeproctext]{ref-wang2017data}
\CSLLeftMargin{9. }%
\CSLRightInline{Wang X, Rush C, Horton NJ. Data visualization on day
one: Bringing big ideas into intro stats early and often. arXiv preprint
arXiv:170508544. 2017. }

\bibitem[\citeproctext]{ref-robinson2017tidyverse}
\CSLLeftMargin{10. }%
\CSLRightInline{Robinson D. Introduction to the tidyverse. DataCamp
Online Course; 2017. Available:
\url{https://www.datacamp.com/courses/introduction-to-the-tidyverse}}

\bibitem[\citeproctext]{ref-robinson2017announcing}
\CSLLeftMargin{11. }%
\CSLRightInline{Robinson D. Announcing "introduction to the tidyverse",
my new DataCamp course. Variance Explained (blog); 2017. Available:
\url{http://varianceexplained.org/r/intro-tidyverse/}}

\bibitem[\citeproctext]{ref-bryan2017twitter}
\CSLLeftMargin{12. }%
\CSLRightInline{Bryan J. Tweet about not front-loading boring
foundational material. Twitter; 2017. }

\bibitem[\citeproctext]{ref-fisher2021better}
\CSLLeftMargin{13. }%
\CSLRightInline{Fisher D, Frey N. Better learning through structured
teaching: A framework for the gradual release of responsibility. ASCD;
2021. }

\bibitem[\citeproctext]{ref-nederbragt2020ten}
\CSLLeftMargin{14. }%
\CSLRightInline{Nederbragt A, Harris RM, Hill AP, Wilson G. Ten quick
tips for teaching with participatory live coding. PLOS Computational
Biology. 2020;16: e1008090. }

\bibitem[\citeproctext]{ref-ebbinghaus1913grundzuge}
\CSLLeftMargin{15. }%
\CSLRightInline{Ebbinghaus H. Grundzüge der psychologie v. 2, 1913.
Veit; 1913. }

\bibitem[\citeproctext]{ref-harris1943habituatory}
\CSLLeftMargin{16. }%
\CSLRightInline{Harris JD. Habituatory response decrement in the intact
organism. Psychological bulletin. 1943;40: 385. }

\bibitem[\citeproctext]{ref-shaw1986donald}
\CSLLeftMargin{17. }%
\CSLRightInline{Shaw G. Donald hebb: The organization of behavior. Brain
theory: Proceedings of the first trieste meeting on brain theory,
october 1--4, 1984. Springer; 1986. pp. 231--233. }

\bibitem[\citeproctext]{ref-swirl}
\CSLLeftMargin{18. }%
\CSLRightInline{Carchedi N, Kross S, Bauer B, et al. Swirl: Learn r, in
r. 2023. Available: \url{https://swirlstats.com}}

\bibitem[\citeproctext]{ref-kaggle_learn}
\CSLLeftMargin{19. }%
\CSLRightInline{Kaggle. Kaggle learn. Online learning platform; 2018.
Available: \url{https://www.kaggle.com/learn}}

\bibitem[\citeproctext]{ref-wickham2023}
\CSLLeftMargin{20. }%
\CSLRightInline{Wickham H, Grolemund G, Çetinkaya-Rundel M. R for data
science. 2nd ed. O'Reilly Media; 2023. }

\bibitem[\citeproctext]{ref-wilson2019teaching}
\CSLLeftMargin{21. }%
\CSLRightInline{Wilson G. Teaching tech together: How to make your
lessons work and build a teaching community around them. Chapman;
Hall/CRC; 2019. }

\bibitem[\citeproctext]{ref-learnr}
\CSLLeftMargin{22. }%
\CSLRightInline{Kross S, Çetinkaya-Rundel M, et al. Learnr: Interactive
tutorials for r. 2024. Available:
\url{https://rstudio.github.io/learnr/}}

\bibitem[\citeproctext]{ref-nbgrader}
\CSLLeftMargin{23. }%
\CSLRightInline{Hamrick JB et al. Nbgrader: A tool for creating and
grading assignments in the jupyter notebook. Proceedings of the 19th
python in science conference. 2016. pp. 68--74.
doi:\href{https://doi.org/10.25080/Majora-629e541a-00e}{10.25080/Majora-629e541a-00e}}

\bibitem[\citeproctext]{ref-otter_grader}
\CSLLeftMargin{24. }%
\CSLRightInline{Kim EJ, Lau S, Hug J, DeNero J. Otter: A tool for
automated grading of jupyter notebooks and more. Proceedings of the 23rd
python in science conference. 2022. pp. 120--127.
doi:\href{https://doi.org/10.25080/majora-2127ed6e-00c}{10.25080/majora-2127ed6e-00c}}

\bibitem[\citeproctext]{ref-horst2020palmerpenguins}
\CSLLeftMargin{25. }%
\CSLRightInline{Horst A, Hill A, Gorman K. Palmerpenguins: Data for
palmer archipelago (antarctica) penguin species. R Journal. 2020;12:
277--283.
doi:\href{https://doi.org/10.32614/RJ-2020-043}{10.32614/RJ-2020-043}}

\bibitem[\citeproctext]{ref-bryan2015stat545}
\CSLLeftMargin{26. }%
\CSLRightInline{Bryan J. STAT 545: Data wrangling, exploration, and
analysis with {R}. 2015. Available: \url{https://stat545.com/}}

\bibitem[\citeproctext]{ref-dplyr}
\CSLLeftMargin{27. }%
\CSLRightInline{Wickham H, François R, Henry L, Müller K. Dplyr: A
grammar of data manipulation. 2024. Available:
\url{https://CRAN.R-project.org/package=dplyr}}

\bibitem[\citeproctext]{ref-canlang}
\CSLLeftMargin{28. }%
\CSLRightInline{Timbers T. Canlang: Canadian census language data. 2020.
Available: \url{https://ttimbers.github.io/canlang/}}

\bibitem[\citeproctext]{ref-gapminder}
\CSLLeftMargin{29. }%
\CSLRightInline{Bryan J. Gapminder: Data from gapminder. 2017.
Available: \url{https://CRAN.R-project.org/package=gapminder}}

\bibitem[\citeproctext]{ref-robinson2021unvotes}
\CSLLeftMargin{30. }%
\CSLRightInline{Robinson D. Unvotes: United nations general assembly
voting data. 2021.
doi:\href{https://doi.org/10.32614/CRAN.package.unvotes}{10.32614/CRAN.package.unvotes}}

\bibitem[\citeproctext]{ref-ihaka1996r}
\CSLLeftMargin{31. }%
\CSLRightInline{Ihaka R, Gentleman R. R: A language for data analysis
and graphics. Journal of computational and graphical statistics. 1996;5:
299--314. }

\bibitem[\citeproctext]{ref-tidyverse}
\CSLLeftMargin{32. }%
\CSLRightInline{Wickham H, Hester J, Henry L, et al. Tidyverse: Easily
install and load the 'tidyverse'. 2024. Available:
\url{https://CRAN.R-project.org/package=tidyverse}}

\bibitem[\citeproctext]{ref-fay2018assignment}
\CSLLeftMargin{33. }%
\CSLRightInline{Fay C. Why do we use arrow as an assignment operator?
Blog post; 2018. Available: \url{https://colinfay.me/r-assignment/}}

\bibitem[\citeproctext]{ref-peng2016r}
\CSLLeftMargin{34. }%
\CSLRightInline{Peng RD. R programming for data science. Leanpub
Victoria, BC, Canada; 2016. }

\bibitem[\citeproctext]{ref-edmondson1999psychological}
\CSLLeftMargin{35. }%
\CSLRightInline{Edmondson A. Psychological safety and learning behavior
in work teams. Administrative science quarterly. 1999;44: 350--383. }

\bibitem[\citeproctext]{ref-lyman2021pre}
\CSLLeftMargin{36. }%
\CSLRightInline{Lyman B, Mendon CR. Pre-licensure nursing students'
experiences of psychological safety: A qualitative descriptive study.
Nurse education today. 2021;105: 105026. }

\bibitem[\citeproctext]{ref-topping1998}
\CSLLeftMargin{37. }%
\CSLRightInline{Topping K. Peer assessment between students in colleges
and universities. Review of Educational Research. 1998;68: 249--276.
Available: \url{http://www.jstor.org/stable/1170598}}

\bibitem[\citeproctext]{ref-gawande2010checklist}
\CSLLeftMargin{38. }%
\CSLRightInline{Gawande A. Checklist manifesto, the (HB). Penguin Books
India; 2010. }

\bibitem[\citeproctext]{ref-parker2018empowering}
\CSLLeftMargin{39. }%
\CSLRightInline{Parker TH, Griffith SC, Bronstein JL, Fidler F, Foster
S, Fraser H, et al. Empowering peer reviewers with a checklist to
improve transparency. Nature ecology \& evolution. 2018;2: 929--935. }

\bibitem[\citeproctext]{ref-ccetinkaya20225ws}
\CSLLeftMargin{40. }%
\CSLRightInline{Çetinkaya-Rundel M, Dogucu M, Rummerfield W. The 5Ws and
1H of term projects in the introductory data science classroom.
Statistics Education Research Journal. 2022;21: 4--4. }

\end{CSLReferences}
\end{document}